\documentclass[footinbib,a4paper,aps,superscriptaddress,reprint,twocolumn,preprintnumbers,amsmath,amssymb,nobalancelastpage,longbibliography,prl]{revtex4-2}

\usepackage[english]{babel}
\usepackage[dvipsnames]{xcolor}
\usepackage{comment}
\usepackage{graphicx}
\usepackage{stackengine}
\usepackage{dcolumn}
\usepackage{bm}
\usepackage[utf8]{inputenc}
\usepackage{amssymb}
\usepackage{amsmath}
\usepackage{pstricks}
\usepackage{slashed}
\usepackage{braket}
\usepackage{amsfonts}
\usepackage{bbm}
\usepackage{physics}
\usepackage{MnSymbol}

\usepackage{natbib}
\usepackage{verbatim}
\usepackage{siunitx}
\usepackage{lipsum}
\usepackage{slashed}
\usepackage{ulem}
\usepackage[left=1.85cm,right=1.85cm,top=1.85cm,bottom=1.85cm]{geometry}
\usepackage{verbatim}
\usepackage{xcolor}
\usepackage{hyperref}
\hypersetup{    colorlinks,    linkcolor={blue!50!black},    citecolor={blue!50!black},    urlcolor={blue!80!black}}

\usepackage{tikz}

\newcommand{\circledsub}[1]{\tikz[baseline=(char.base)]{\node[shape=circle,draw,inner sep=0.5pt,scale=0.9] (char) {#1};}}
\newcommand{\circledsup}[1]{\tikz[baseline=(char.base)]{\node[shape=circle,draw,inner sep=0.5pt,scale=0.5] (char) {#1};}}

\makeatletter
\renewcommand{\p@paragraph}{} 
\makeatother
\usepackage{lipsum}

\begin{document}

\title{Programmable Fermionic Quantum Processors with Globally Controlled Lattices}

\author{Gabriele Calliari}\thanks{These authors contributed equally to this work.}
\author{Charles Fromonteil}\thanks{These authors contributed equally to this work.}
\author{Francesco Cesa}
\author{Torsten~V.~Zache}
\affiliation{Institute for Theoretical Physics, University of Innsbruck, Innsbruck, 6020, Austria}
\affiliation{Institute for Quantum Optics and Quantum Information, Austrian Academy of Sciences, Innsbruck, 6020, Austria}

\author{Philipp M. Preiss}
\affiliation{Max Planck Institute of Quantum Optics, Hans-Kopfermann-Str.1, Garching D-85748, Germany}
\affiliation{Munich Center for Quantum Science and Technology (MCQST), Schellingstr. 4, M\"unchen D-80799, Germany}

\author{Robert Ott}
\author{Hannes Pichler}\email{Hannes.Pichler@uibk.ac.at}
\affiliation{Institute for Theoretical Physics, University of Innsbruck, Innsbruck, 6020, Austria}
\affiliation{Institute for Quantum Optics and Quantum Information, Austrian Academy of Sciences, Innsbruck, 6020, Austria}

\begin{abstract}
We introduce a framework for realizing universal fermionic quantum processing with globally controlled itinerant fermionic particles.
Our approach is tailored to the example of neutral atoms in optical lattices, but transposes to other setups with similar capabilities.
We give constructive protocols to realize arbitrary fermionic processes, with time-dependent control over global parameters of the experimental setup, such as tunneling and interaction in a Fermi-Hubbard type model. 
We first prove the universality of our framework and then discuss implementation variants, such as hybrid analog-digital simulation of extended Fermi-Hubbard models, e.g., with long-range couplings.

\end{abstract}

\maketitle

\textit{Introduction---}Understanding fermionic many-body physics is a central topic in quantum science, bridging chemistry, condensed matter, high-energy physics, and cosmology. 
However, calculating properties of interacting systems of fermions with classical computers is, in general, prohibitively difficult even for moderate system sizes~\cite{troyer2005computational}.
In recent years, there have been growing efforts to address this challenge with quantum simulators~\cite{altman2021quantum,daley2022practical} based on itinerant fermionic particles, where fermionic statistics are naturally enforced at the physical level. For instance, neutral atoms in optical lattices~\cite{gross2021quantum,schafer2020tools}
or electrons in semiconductor quantum dot arrays~\cite{hanson2007spins,hensgens2017quantum,boter2022spiderweb,kunne2024spinbus,donnelly2026largescale} naturally give rise to scalable simulations of paradigmatic fermionic models, most notably the Fermi-Hubbard model with nearest-neighbor tunneling and on-site interaction.

An outstanding challenge is to access models that are not natively realized in these platforms and require a larger degree of programmability. Often, experimental control is in fact limited to a few, \textit{uniformly} tunable parameters of the microscopic Hamiltonian. This motivates the exploration of the degree of programmability that can be achieved under these constraints. While similar questions have been object of intense research for qubit-based setups,~\cite{lloyd1993potentially,cirac2000scalable,benjamin2000schemes,calarco2004quantum,raussendorf2005quantum,shepherd2006universally, lloyd2016adiabatic, wintermantel2020unitary,cesa2023universal,PhysRevResearch.7.L012065, hu2026universal}, fermionic processing features intrinsically different native operations~\cite{bravyi2002fermionic}, which are also subject to fundamental constraints, such as superselection rules; in circuit formalism, this results in different gate sets. Thus, fermionic quantum processing protocols with global control require new strategies.

Here we introduce a framework for fermionic quantum processing under such conditions. Our approach is designed for neutral atoms in optical lattices, and assumes only time-dependent control over a superlattice potential, a magnetic field gradient, and the contact interaction strength. We provide explicit protocols that harness these tools to realize a universal fermionic gate set, enabling the construction of arbitrary quantum processes in a programmable way. Due to its global control character, the scheme is highly parallelizable in large one- and two-dimensional lattices, and thus well-suited for quantum simulation tasks.

We first provide a constructive proof of universality in a minimal setup consisting of a one-dimensional (1D) lattice with spin-1/2 fermions: atoms in one spin state are used as a fermionic register, while a single atom in the other spin state acts as a control particle. We show how the latter can be moved through the lattice via global sequences, and used to trigger operations in a sequential way akin to a Turing head, similar to Refs.~\cite{cirac2000scalable,calarco2004quantum}. The scheme requires local programmability once to initialize the control atom in a known location (see, e.g., Refs.~\cite{weitenberg2011single,preiss2015strongly,young2022tweezer}), when the data register is in a product state. During processing, when fragile quantum coherences are present, our protocol relies solely on global control sequences.

After proving universality, we consider variants of such a construction, e.g., tailored towards quantum simulation tasks -- which, in our framework, are considerably simpler than full universal processing. 
For example, we discuss parallel operation on multiple copies of the same state or faster implementations of structured circuits using multiple control heads.
In particular, we discuss how to exploit 1D and 2D lattice setups to implement digital Trotterizations of fermionic dynamics. 
Finally, we combine this approach with analog evolution under the native Hamiltonian, enabling hybrid analog-digital simulation~\cite{tabares2025programming} of extended Fermi-Hubbard models~\cite{schlomer2024local}.  

\begin{figure*}[ht]
    \centering
    \includegraphics[width=0.9\linewidth]{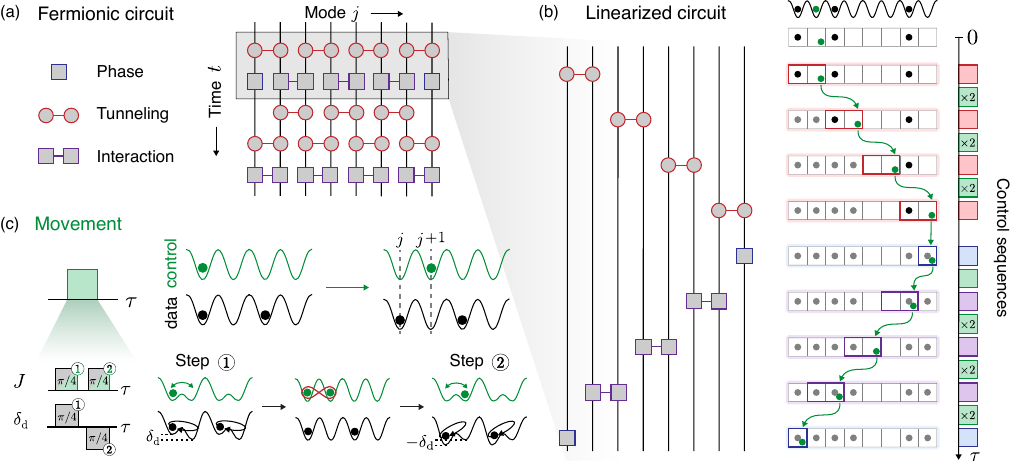}
\caption{\textbf{Architecture}.
(a) An example fermionic quantum circuit, composed of local phase, tunneling and interaction gates. 
(b)
We realize the quantum circuit in a 1D optical lattice by placing a \textit{control} fermion (green dots) in the target location of the gate; for two-mode gates we place it in the right well of the corresponding double well (DW). With a single control fermion we can implement linearized circuits, i.e., we allow exactly one gate per circuit layer. Circuits are applied by alternating control fermion movement with tailored global control sequences acting non-trivially on data atoms (black dots) only in the DW where the control fermion sits, see main text and Fig.~\ref{fig:Fig2}. (c) A key element is the selective movement of the control atoms. While data fermions are fixed at a given spatial mode $j$, the control atom is moved as follows (green shaded box): First, the lattice is rearranged into DWs along axis~$j$, such that the control atom sits on a left well~L. We then allow tunneling with strength $J$ for time $\tau=\pi/(4J)$, in presence of a gradient $\delta_\mathrm{d}=\delta_\mathrm{\uparrow}$ for data fermions. This step is repeated with inverted gradient $-\delta_\mathrm{d}$. By appropriately choosing the value $\delta_\mathrm{d}$, data fermions return to their original wells in each step, while the control fermion proceeds by one site.
}
    \label{fig:Fig1}
\end{figure*}

\textit{Setup---}We first describe a 1D setup for universal quantum processing, and generalize to other settings further below. We consider a set of fermionic modes with annihilation (creation) operators $f^{(\dagger)}_{\sigma,j}$, labeled by a 1D spatial mode index $j\in\{0,1,..,L-1\}$
and an internal-state (spin) index $\sigma \in\{\uparrow, \downarrow\}$, which fulfill canonical anti--commutation relations $\{f^\dagger_{\sigma,j},f_{\sigma',j'}\}=\delta_{\sigma\sigma'}\delta_{jj'}$. We assume that the microscopic dynamics are described by the Hamiltonian
\begin{equation}
        H_\mathrm{sys}=H_{J} + H_\delta + H_U.
    \label{eq:Hamiltonian}
\end{equation}
The three terms represent nearest-neighbor tunneling in a staggered lattice, 
spin-dependent energy gradient, and on-site interactions, respectively; this Hamiltonian is realized, e.g., by neutral atoms in optical superlattices.

The first term, $H_J = H_{J_\mathrm{E}}+H_{J_\mathrm{O}}$, with
\begin{equation}
    H_{J_\mathrm{B}} =  \textstyle - J_\mathrm{B} \sum_{\sigma,j\in \mathrm{B}} (f^\dagger_{\sigma,j}f_{\sigma,j+1} + \text{H.c.})\\
    \label{eq:HamTunn}
\end{equation} 
(where $\mathrm{B}\in\{\mathrm{E},\mathrm{O}\}$ denotes even or odd links), describes particle tunneling with strength $J_\mathrm{B}$ between neighboring sites in the staggered lattice~\cite{mandel2003controlled,folling2007direct,cheinet2008counting,trotzky2008timeresolved}. These tunneling strengths can be tuned with respect to each other by adjusting the lattice parameters.

The second term introduces a spin-dependent energy offset between neighboring sites, e.g., with a magnetic field gradient described by
\begin{equation}
\textstyle
    \label{eq:HamDetuning}
    H_\delta  =  \sum_{\sigma,j} j\delta_{\sigma} \, n_{\sigma,j},
\end{equation}
with number operator $n_{\sigma,j}=f_{\sigma,j}^\dagger f_{\sigma,j}$ and spin-dependent $\delta_{\sigma}$~\footnote{Instead of independently controlling both $\delta_{\sigma}$, $H_\delta$ can be realized by tuning the differential gradient $\delta_{\uparrow}-\delta_{\downarrow}$ (e.g., using a magnetic field gradient) and the common-mode one $\delta_{\uparrow}+\delta_{\downarrow}$ (e.g., by tilting the optical lattice).}.

Finally, the third term
 \begin{equation}
 \textstyle
     H_U = U\sum_{j}n_{\uparrow,j}n_{\downarrow,j}
     \label{eq:HamInteraction}
 \end{equation}
describes the contact interaction with strength $U$ between opposite spin particles occupying the same site, which can be tuned, e.g., via Feshbach resonances~\cite{chin2010feshbach}, or by switching between interacting and non-interacting internal states. 

In summary, we have five independent global control parameters $J_\mathrm{E}, J_\mathrm{O}, \delta_\uparrow, \delta_\downarrow, U$ that can be tuned time-dependently. Our protocols below are all based on global control sequences of these parameters.

A central ingredient in our framework is the possibility to pairwise isolate the lattice sites in double wells (DWs) on even (odd) bonds, by setting $J_\mathrm{O}=0$ ($J_\mathrm{E} =0$).
Within such isolated DWs, $H_J$ induces tunneling between left and right wells, $H_\delta$ generates a potential offset between them, and $H_U$ can be exploited to implement non-Gaussian operations. 
Combined, these ingredients are sufficient for 
universal control of DW states~\cite{mark2025efficiently}; such controlled manipulations in DWs have also been the focus of recent experiments in atomic platforms~\cite{impertro2025strongly,zhu2025splitting,kiefer2026protected,bojovic2026highfidelity}.

\begin{figure*}
    \centering
\includegraphics[width=\linewidth]{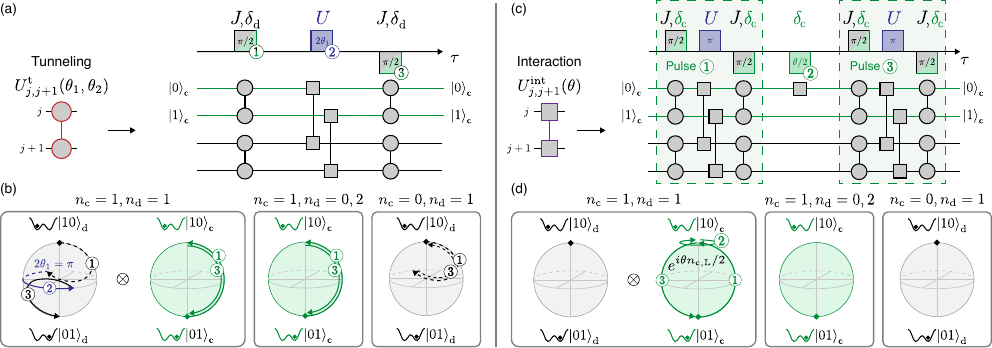}
    \caption{\textbf{Gate sequences.} Implementation of tunneling [(a)-(b)] and interaction gates [(c)-(d)].
    (a) Tunneling gate for data modes (black) in presence of a control fermion (green) in the same DW. The gate is decomposed into a sequence of physical operations on the atoms: tunneling (black circles) and contact interaction (black squares). We realize these operations using time-dependent global control of tunneling strength $J$ (for both data and control), data gradient $\delta_\mathrm{d}$, and interaction~$U$. This circuit implements $U_{j,j+1}^\mathrm{t}(\theta_1,\theta_2\approx2.4959)$, but can be generalized to arbitrary $\theta_2$~(see SM). (c) Interaction gate. In the first step, we re-use the gate sequence developed for the tunneling gate, with exchanged roles of data and control fermions: this brings the control fermion to the left DW site, conditioned on the data fermion parity within the DW. The second step applies a differential phase (single black square) to control fermions in the left DW site. Third, we repeat the first step. (b),(d) Bloch sphere representations of the state trajectories for different example configurations $\ket{n_\mathrm{c,L}n_\mathrm{c,R}}_\mathrm{c}$ and $\ket{n_\mathrm{d,L}n_\mathrm{d,R}}_\mathrm{d}$ of control and data atoms.}
    \label{fig:Fig2}
\end{figure*}

Our goal is the implementation of an arbitrary fermionic circuit acting on $L$ modes with fixed particle number $N\leq L$; this can be decomposed into single- and two-mode fermionic gates~\cite{bravyi2002fermionic}, drawn, e.g., from the universal set 
\begin{align}\label{eq:gateset}
\mathcal{G}=\big\{& U^\mathrm{p}_j (\theta)=e^{-i\theta n_{\mathrm{d},j}}, U^\mathrm{int}_{j,j+1}(\theta)=e^{-i\theta n_{\mathrm{d},j}n_{\mathrm{d},j+1}},\nonumber\\
& U^{\mathrm{t}}_{j,j+1}(\theta_1,\theta_2)=e^{-i\theta_1 (e^{-i\theta_2}d_{j}^\dagger d_{j+1} + \rm {H.c.})}\big\}\, ,
\end{align}
where $\left\{d^\dag_i,d_j\right\}=\delta_{ij}$ and $n_{\mathrm{d},j}=d^\dag_j d_j$. This gate set comprises the single-mode phase gate $U^\mathrm{p}$, and two two-mode gates: the interaction gate $U^\mathrm{int}$ and the tunneling gate $U^\mathrm{t}$ [see Fig.~\ref{fig:Fig1}(a)].  

In our approach, we identify fermions with spin $\sigma=~\uparrow$ as \textit{data} particles, i.e., $d_j \equiv f_{\uparrow, j}$.  
Additionally, we employ one particle of the opposite spin state, $\sigma=\,\downarrow$, as a \textit{control} particle, with $c_j\equiv f_{\downarrow, j}$, which can trigger local quantum gates on data modes similar to a Turing head~\cite{calarco2004quantum}. 
Accordingly, in the following, we relabel the mode index $\sigma$ from $\sigma\in\{\mathrm{\uparrow},\mathrm{\downarrow}\}$ to $\sigma\in\{\mathrm{d},\mathrm{c}\}$.

Specifically, to implement gates we move the control atom to the target location~\cite{cirac2000scalable,calarco2004quantum} via global control sequences, and exploit the data-control interaction to trigger the local gate~\cite{jaksch1999entanglement}, see Fig.~\ref{fig:Fig1}(b). 
For single-mode gates, the presence of the control atom directly enables the implementation of the phase gate.
For two-mode gates, we shape the superlattice into an array of isolated DWs, and apply a global control sequence that realizes the desired gate at the DW hosting the control atom, while leaving all the other DWs unaffected.
In the following, we provide detailed protocols for both the movement of the control atom, and the implementation of quantum gates. 
Related protocols have been the focus of recent works in the context of locally controlled fermionic quantum processing~\cite{gonzalez-cuadra2023fermionic,gkritsis2025simulating,roth2026constructing}, or globally controlled measurements of nearest-neighbor observables~\cite{mark2025efficiently}; 
here, we introduce protocols to \emph{locally} implement a universal gate set using only \emph{global} control.

\textit{Control atom movement \& Gates---}We first present a protocol to move the control atom without affecting data atoms. This selective movement can be realized in different ways, for instance, via atom shuttling with spin-dependent lattices
~\cite{jaksch1999entanglement,mandel2003coherent,gonzalez-cuadra2023fermionic}, 
or via high-fidelity topological pumping~\cite{citro2023thouless}. Here, we present a protocol which only requires temporal control over the parameters defined in Eqs.~\eqref{eq:Hamiltonian}-\eqref{eq:HamInteraction}. To this end, we first deform the lattice into DWs, and allow tunneling for a total time $\tau=\pi/(2J)$ within each DW: as a consequence, the control atom moves from one site of the DW to the other. By repeating this procedure with DWs on odd and even bonds of the lattice, the atom can be moved deterministically to arbitrary locations.
The tunneling of the data atoms is inhibited using a simultaneous spin-dependent energy offset $\delta_\mathrm{d}$. 
Specific values of $\delta_\mathrm{d}$ ensure that data atoms exactly return to their initial position after each step, 
as shown in Fig.~\ref{fig:Fig1}(c) and detailed in the End Matter (EM)~\footnote{By evolving with a non-zero gradient $\delta_\mathrm{d}$, the state acquires single-mode phases. These can be avoided, e.g., by splitting the evolution in two steps with opposite gradient.};
alternatively, setting $\delta_\mathrm{d} \gg J$ strongly suppresses tunneling by introducing a large energy difference between neighboring sites.

Once the control atom has been positioned at the target site, the gates of $\mathcal{G}$ can be triggered by the following sequences.

\textit{(i) Phase gate.} With the control atom in site $j$, a collisional phase gate $U_{j}^\mathrm{p} (\theta)$ can be realized by turning on the interaction $U$ for a physical time $\tau = \theta/U$~\cite{jaksch1999entanglement,brennen1999quantum,mandel2003coherent}. This implements, at each site $i$, the unitary $\mathrm{exp}[-i \theta n_{\mathrm{c},i} n_{\mathrm{d},i}]$, with $n_{\mathrm{c},i}=c_i^\dagger c_i$. At site $j$ ($n_{\mathrm{c},j}=1$), this unitary imprints a single-mode phase $\theta$, while for other sites ($n_{\mathrm{c},i}=0$) it reduces to the identity. Thus, the protocol implements $U_{j}^\mathrm{p}(\theta)$ on the data modes.

\textit{(ii) Tunneling gate.}  
We now describe the implementation of $U^\mathrm{t}_{j,j+1}(\theta_1,\theta_2)$ with the control atom now sitting in site $j+1$.
First, we pairwise isolate the lattice sites such that $j$ and $j+1$ are the left and right wells of a DW~\footnote{The gate can be implemented in a similar way with a control atom in the left well, see Supplemental Material}.
We then perform a three-step control sequence, as shown in Fig.~\ref{fig:Fig2}(a). First \circledsub{1}, we allow tunneling for a duration $\tau = \pi/(2J)$, with a nonzero offset $\delta_\mathrm{d}\approx 1.5974 J$ on the data fermions: this sends the control atom from the right DW site to the left, i.e., site $j$, and makes the data atoms tunnel non-trivially. Second \circledsub{2}, we turn on the interaction $U$ for time $\tau=2\theta_1/U$, which imprints a phase on the $j$ data mode.
Finally \circledsub{3}, we invert the first step (i.e., $J,\delta_\mathrm{d}\rightarrow-J,-\delta_\mathrm{d}$): the control atom returns to its initial position, and the data atoms tunnel non-trivially again~\footnote{We can invert the tunneling strength $J\rightarrow -J$ with extra relative phases, implemented by an additional control sequence, see Supplemental Material.}.

The action of the above sequence can be understood as follows. The data modes in a DW live in a four-dimensional Hilbert space spanned by the Fock states $\ket{n_\mathrm{d,L}n_\mathrm{d,R}}_\mathrm{d}$. The sequence acts non-trivially only on the two-level subspace $\mathrm{span}\{\ket{01}_\mathrm{d},\ket{10}_\mathrm{d}\}$, for which tunneling and interaction correspond to Bloch sphere rotations [see Fig.~\ref{fig:Fig2}(b)]. In this representation, steps \circledsub{1} and \circledsub{3} amount to a basis change which maps the $z$-axis onto the equatorial plane, and step \circledsub{2} implements a rotation of angle $2\theta_1$ around the $z$-axis, conditioned on the presence of the control atom. Thus, in DWs with no control atom steps \circledsub{1} and \circledsub{3} simply cancel each other out, while in the DW $(j,j+1)$, hosting the control atom, the data atoms tunnel to realize $U^\mathrm{t}_{j,j+1}(\theta_1,\theta_2)$ up to single-mode phases~\footnote{These phases can be actively corrected using additional phase gates, or kept track of when compiling the circuit.}. 
The parameter $\theta_2$ is fixed by the control sequence, but can be arbitrarily changed by turning on gradients $\delta_\mathrm{d}$ before and after the sequence [see Supplemental Material (SM)].

\textit{(iii) Interaction gate.} Similarly, with the control atom in site $j+1$, we implement the interaction gate~$U_{j,j+1}^{\mathrm{int}}(\theta)$ as follows [see Fig.~\ref{fig:Fig2}(c-d)]. We arrange the lattice into DWs in the same way as for the tunneling gate, and then we apply the following sequence: \circledsub{1} We implement a variant of the tunneling gate control sequence with $\theta_1=\pi/2$, where we exchange the role of data and control atoms ($\delta_\mathrm{d}\leftrightarrow\delta_\mathrm{c}$): this moves the control atom to the left site if there is exactly one data atom in its DW.
\circledsub{2} We switch on an energy offset $\delta_\mathrm{c}$ on control atoms for a time $\tau=-\theta/(2\delta_\mathrm{c})$, to imprint a phase on control atoms in the left wells. \circledsub{3} We repeat step \circledsub{1}.

The crucial part of the sequence is step \circledsub{2}, which imprints a phase on the control atom if it has moved in step \circledsub{1}, i.e., only for odd-parity data configurations; then,
step \circledsub{3} returns the control atom to its original site. 
Therefore, in the DW $(j,j+1)$ the data basis states $\ket{01}_\mathrm{d}$ and $\ket{10}_\mathrm{d}$ acquire a phase $e^{i\theta/2}$, while $\ket{00}_\mathrm{d}$ and $\ket{11}_\mathrm{d}$ are unaffected. This implements the unitary $U^{\mathrm{int}}_{j,j+1}(\theta)$ up to a single-mode phase $\mathrm{exp}[i(\theta/2)(n_{\mathrm{d},j}+n_{\mathrm{d},j+1})]$.
We provide additional details in the SM.

\textit{Generalizations---}So far, we have demonstrated universal fermionic processing in a 1D lattice setup with a single control fermion. In the following, we discuss example extensions of this architecture. 

\begin{figure}[t]
    \centering
    \includegraphics[width=0.98\linewidth]{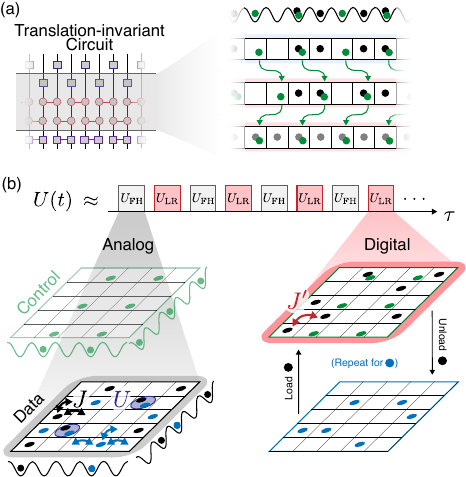}
    \caption{\textbf{Hamiltonian simulation}. 
    (a)~The computation is parallelized with multiple control heads, to simultaneously implement gates in multiple locations. This is particularly useful for translation-invariant circuits, e.g., for Hamiltonian simulations of fermionic models.
    (b) Trotter circuits for extended Fermi-Hubbard models can be composed by alternating steps of short analog evolution~$U_{\mathrm{FH}}$ [with native Hamiltonian Eq.~\eqref{eq:Hamiltonian}] and, e.g., digital long-range tunneling gates~$U_{\mathrm{LR}}$ performed in a second optical lattice plane. Here, it suffices to initialize the control heads in the shown four-fold symmetric pattern (see EM). Black and blue dots represent spin-$\uparrow$ and -$\downarrow$ fermions, green dots represent control particles, e.g., of a third atomic spin state.
    }
    \label{fig:Fig3}
\end{figure}

For instance, for sufficiently structured quantum circuits (e.g., translation-invariant Trotter circuits) our protocol can be optimized by using multiple control atoms as control heads, placed in the lattice according to spatial patterns reflecting the circuit symmetries, see Fig.~\ref{fig:Fig3}(a). In this way, each control sequence simultaneously implements the same gate at multiple lattice sites. Additionally, our protocol can be trivially parallelized by operating simultaneously on multiple copies of a quantum state in a single optical lattice. All our constructions straightforwardly extend to higher dimensions: by loading atoms in 2D optical lattices and moving the control atoms with superlattices in both spatial directions, we can implement arbitrary fermionic circuits with 2D gate connectivity, see SM.

In addition, our setup enables hybrid analog-digital fermionic quantum processing, taking advantage of both the native evolution and our universal gate set to explore complex models.
We illustrate this with an extension of the 2D Fermi-Hubbard Hamiltonian~$H_{\mathrm{FH}} = \textstyle -\sum_{\langle ij\rangle,\sigma} J(c^\dagger_{\sigma,i}c_{\sigma,j} + \mathrm{H.c.}) + U\sum_i n_{\uparrow,i}n_{\downarrow,i}$ (where $\langle \cdot \rangle $ denotes nearest-neighbor lattice sites). While $H_{\mathrm{FH}}$ is natively included in the Hamiltonian Eq.~\eqref{eq:Hamiltonian}, extensions can be added using the digital gate operations developed above: an example is  $H_{\mathrm{LR}}  = - \sum_{\llangle ij\rrangle,\sigma} J'(c^\dagger_{\sigma,i}c_{\sigma,j} + \mathrm{H.c.})$, representing long-range hopping between next-nearest-neighboring lattice sites ($\llangle \cdot \rrangle$). 

This hybrid analog--digital operation mode is illustrated in Fig.~\ref{fig:Fig3}(b): We first prepare both spin-$\uparrow$ and $\downarrow$ particles in a ``data'' plane of the 2D lattice. Second, we place additional control atoms, e.g., in a third spin state~\cite{mongkolkiattichai2025quantum}, in a separate ``control'' plane with a four-fold configuration pattern, see Fig.~\ref{fig:Fig3}(b). For the evolution, we split the unitary time-evolution operator $U(t) = \exp[-i(H_{\mathrm{FH}}+H_{\mathrm{LR}})t]$ into Trotter steps given by $U_{\mathrm{FH}} =  \exp(-iH_\mathrm{FH}\Delta t)$ and $U_\mathrm{LR} = \prod_{\llangle ij\rrangle,\sigma} \exp[iJ'\Delta t(c^\dagger_{\sigma,i}c_{\sigma,j} + \mathrm{H.c.})]$, such that $U(t) = (U_{\mathrm{FH}}U_\mathrm{LR})^{t/\Delta t}+\mathcal{O}(t\Delta t)$. Here, $U_{\mathrm{FH}}$ represents the analog evolution with the native Hamiltonian Eq.~\eqref{eq:Hamiltonian} (with $J_\mathrm{E}=J_\mathrm{O}$, and $\delta_\sigma = 0$), which acts on the data particles. Conversely, $U_\mathrm{LR}$ is implemented with a sequence of digital operations: \circledsub{1} We use the movement protocol to move all $\uparrow$-fermions from the data to the control plane. \circledsub{2}~We apply the next-nearest-neighbor tunneling unitary on all $\llangle ij\rrangle$ pairs, via interactions with the control fermions, through a series of gate sequences and control atom movements; 
see details in EM and SM. \circledsub{3}~$\uparrow$-fermions are moved back to the data plane. The whole sequence is then repeated for $\downarrow$-fermions.

To avoid tunneling of control particles during the analog evolution, we choose a setup where control atoms are strongly confined by the lattice. We note that alternative variants of this hybrid analog-digital protocol exist, where no third spin state is required, or, where we perform both analog and digital operations in a single optical lattice plane, see SM for a detailed explanation. Beyond the example discussed in this section, multiple other operation modes are conceivable, e.g., by introducing multiple control planes with different particle patterns to realize more complex circuits, see also~SM.

\textit{Initialization---}For the atom initialization, the different variants of our protocol require different levels of experimental control. For full universality, we require initialization of individual control atoms at \emph{known} lattice locations. For instance, this can be achieved deterministically using additional local optical potentials (see, e.g., Ref.~\cite{weitenberg2011single,preiss2015strongly,young2022tweezer}) used \emph{once} during the initial step. Since this local control is applied before the computation, while the system is typically in a product state, detrimental noise effects on the coherences are avoided.

Conversely, initializing the system to realize translation invariant circuits requires no local control. For instance, to prepare a regular array of control heads, as shown in Fig.~\ref{fig:Fig3}(a), control atoms can be directly initialized with staggered occupation in the optical superlattice, see, e.g., experimental demonstrations in Refs.~\cite{yang2020cooling,chalopin2025optical}.

Depending on the experimental capabilities, other initialization schemes are possible: Alternatively, control particles could be randomly loaded into the lattice during state initialization without requiring any local control. Non-destructive measurements detect the classical location of the control atom \emph{before} the computation and subsequently the control head is again moved to the target location using our protocols. Without such measurement capabilities, one can initialize multiple copies of the state, randomly load the control atoms, measure all atoms \emph{after} the computation, and post-select the results for the cases where the control head was in the correct location.

\textit{Conclusions---}In this article, we introduced a framework for globally controlled fermionic quantum processing by directly manipulating itinerant \emph{fermionic} particles~\cite{gonzalez-cuadra2023fermionic,ott2025errorcorrected,schuckert2025faulttolerant}. 
Our approach achieves universality by applying sequences of uniform adjustments of the Hamiltonian parameters. 
The specific control sequences presented here can be further improved, using for instance optimal control methods~\cite{glaser2015optimalcontrol} to minimize gate duration and susceptibility to noise. Additionally, the protocols can be further tailored to available experimental capabilities, e.g., to avoid fast time-dependent control of the interaction strength or the spin-dependent gradients.

Our scheme is primarily designed for neutral atoms in optical lattices and relies on high-fidelity control over fermionic motional states, as recently demonstrated~\cite{hartke2022quantum,bojovic2026highfidelity,kiefer2026protected}. Its applicability also extends to atoms in reconfigurable tweezer arrays~\cite{kaufman2021quantum,becher2020measurement,yan2022twodimensional,tao2024high,jain2025programmable,eckner2025assembling} and to other platforms, e.g., semiconductor quantum dot arrays~\cite{dehollain2020nagaoka,langrock2023blueprint}.

\textit{Acknowledgments---}We thank Liyuan Chen, Henning Schlömer and Susanne Yelin for discussions and Peter Zoller for discussions and collaboration on related work. This work was funded by the Austrian Science Fund (FWF) through the SFB Qnnect (Grant-DOI 10.55776/F101200), and the COE Quantum Science Austria (Grant No. DOI 10.55776/COE1). This work is supported by the European Union’s Horizon Europe research and innovation program under Grant Agreement No. 101113690 (PASQuanS2.1), the ERC Starting Grant QARA (Grant No. 101041435), and the EU-QUANTERA project TNiSQ (N-6001).
P.M.P acknowledges funding from the German Federal Ministry of Research, Technology and Space (BMFTR grant agreement 13N15890, FermiQP) and the European Union’s Horizon 2020 research
and innovation program (ERC StartingGrant UniRand — Grant No. 948240).
T.V.Z. is supported by an ERC Starting Grant (QS-Gauge, Grant No. 101220401).
Funded by the European Union. Views and opinions expressed are however those of the authors only and do not necessarily reflect those of the European Union or the European Research Council Executive Agency. Neither the European Union nor the granting authority can be held responsible for them.

\bibliographystyle{apsrev4-2_max10authors} 
\bibliography{bibliography}


\appendix

\setcounter{equation}{0}
\setcounter{table}{0}
\makeatletter
\setcounter{subsection}{0}
\renewcommand{\theequation}{A\arabic{equation}}
\let\oldthesubsection\thesubsection
\renewcommand{\thesubsection}{Appendix~\Alph{subsection}}

\onecolumngrid
\section*{ End Matter }
\twocolumngrid

\subsection{Moving the control atoms}

Here, we detail an exact control sequence for the control atom movement, designed to have no net effect on the data fermions~[see Fig.~\ref{fig:Fig1}(c)]. The lattice is deformed into an array of DWs, within which atoms can tunnel with strength $J$. As stated in the main text, we need to allow tunneling for a total time $\tau=\pi/(2J)$ to ensure that the control atom changes site. We divide the sequence in two steps: \circledsub{1} We apply a spin-dependent offset $\delta_\mathrm{d}=2J\sqrt{16m^2-1}$ ($m\in \mathbb{N}$) for data fermions in the left well, and allow tunneling for time $\tau = \pi/(4J)$. \circledsub{2}~We invert the sign of the potential offset ($\delta_\mathrm{d}\rightarrow -\delta_\mathrm{d}$) and repeat step \circledsub{1}.

The choice of $\delta_\mathrm{d}$ ensures that in both steps, data atoms perform $m$ round trips and return to their original site; by inverting the gradient, the single-mode phases they acquire during the evolution cancel out. Conversely, the control atoms, initially placed in the left well, are not affected by $\delta_\mathrm{d}$, and thus tunnel for a total time $\pi/(2J)$, i.e., they fully transition to the subsequent site. The direction of the control atom movement is determined by the initial (even or odd) DW arrangement chosen.

\subsection{Tunneling of control atoms conditioned on data configuration}
In the main text, we discussed how, upon replacing the gradient $\delta_\mathrm{d}$ for data atoms with a gradient $\delta_\mathrm{c}$ for control atoms, the tunneling gate control sequence \circledsub{1}-\circledsub{3} implements a tunneling operation on control fermions conditioned on the initial presence of a data atom in the right well. 
Here, we extend the discussion to different initial data configurations, and describe the action of the sequence on control atoms in those sectors. This is an essential step in order to exploit the sequence as a subroutine for the implementation of the interaction gate. 

We start from the case of no data atoms: the interaction step  \circledsub{2} reduces to the identity, and thus the other steps simply cancel out pairwise. The whole sequence then corresponds to an identity on the control atoms.
Instead, as partially discussed in the main text, in the presence of a single data particle in the DW the control sequence implements the operation $\mathrm{exp}{[-i\theta_1 (n_\mathrm{c,L}+n_\mathrm{c,R})]} \cdot {\mathrm{exp}[- i\theta_1(e^{-i\theta_2} c^\dagger_\mathrm{L} c_\mathrm{R}+\rm H.c.)]}$ on the control atoms, with $\theta_2\approx 2.4959$ (resp. $\theta_2~\approx~\pi+2.4959$) for a data atom in the right (resp. left) DW well, see SM. 
Finally, for two data atoms, the interaction step \circledsub{2} simply implements a phase $\theta_1$ in both wells, while the remaining steps cancel out: the total operation on control atoms is thus $\exp[-2i\theta_1(n_\mathrm{c,L}+n_\mathrm{c,R})]$.

In summary, the tunneling control sequence for $\delta_\mathrm{c}$ implements the following operation:
\begin{align}
&(1-n_\mathrm{d,L})(1-n_\mathrm{d,R})\mathbbm{1} +n_\mathrm{d,L}n_\mathrm{d,R}e^{-2i\theta_1(n_\mathrm{c,L}+n_\mathrm{c,R})} \nonumber \\
&+ (1-n_\mathrm{d,L})n_\mathrm{d,R}
 e^{-i\theta_1 (n_\mathrm{c,L}+n_\mathrm{c,R})}e^{-i\theta_1(e^{-i\theta_2}c_\mathrm{L}^{\dagger}c_\mathrm{R}+\rm H.c.)} \nonumber \\
 &+ (1-n_\mathrm{d,R})n_\mathrm{d,L}
 e^{-i\theta_1 (n_\mathrm{c,L}+n_\mathrm{c,R})}e^{+i\theta_1(e^{-i\theta_2}c_\mathrm{L}^{\dagger}c_\mathrm{R}+\rm H.c.)}. \nonumber
\end{align}

\subsection{Digital implementation of next-nearest-neighbor tunneling}

Here, we describe the digital implementation of the long-range tunneling unitary $U_\mathrm{LR}$ in the hybrid quantum simulation scheme presented in the main text for the extended Fermi-Hubbard model $H_\mathrm{FH}+H_\mathrm{LR}$.

We consider a 2D control plane where only the modes with even row and column indices are filled, see Fig.~\ref{fig:Fig3}(b). Next-nearest-neighbor tunneling is achieved, for each spin species, by loading all atoms of that species to the control plane, and then executing a sequence of digital tunneling gates and lattice rearrangements. Specifically, for the hopping between the bottom left and the top right corner of a four-site square we perform the sequence
\begin{equation}
    \includegraphics[width=0.9\linewidth]{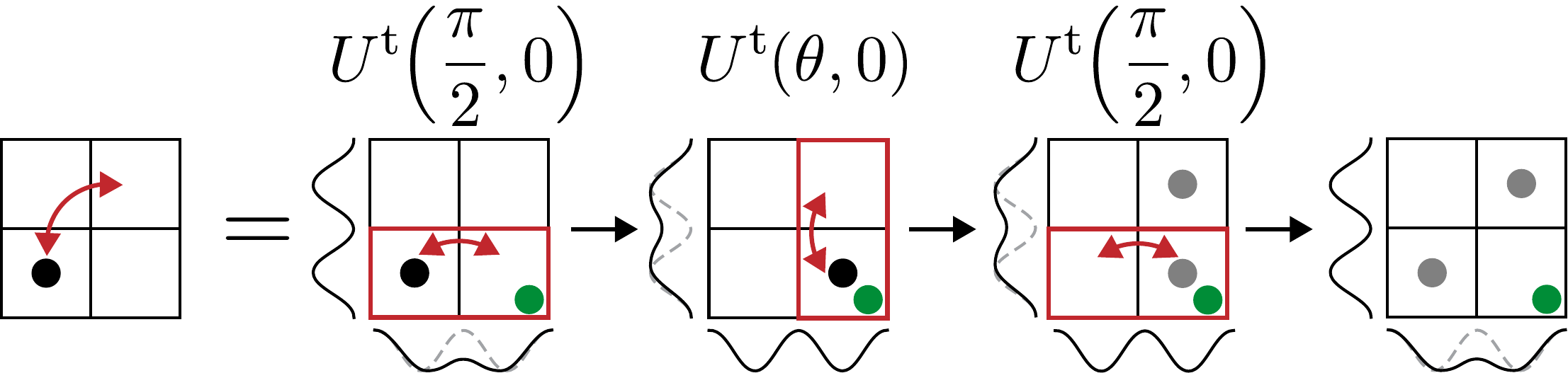}.
\end{equation}
The key idea is to first transport data fermions between the bottom left and the bottom right corner, and then apply the desired tunneling gate between the modes in the right column.
To this end, we first realize a DW structure in the horizontal direction, and apply the control sequence for the tunneling gate $U^\mathrm{t}(\pi/2,0)$, such that the data fermions within the DW exchange their position. We then rearrange the lattice to feature DWs in the vertical direction, and apply the tunneling gate $U^\mathrm{t}(\theta,0)$, with $\theta=-J'\Delta t$. Finally, we rearrange the lattice to again exhibit DWs in the horizontal direction, and reapply the $U^\mathrm{t}(\pi/2,0)$ control sequence to revert the first operation. 

The full unitary $U_{\mathrm{LR}}$ is implemented by repeating the above sequence eight times for each species. 
The diagonal term between top left and bottom right is realized with the same sequence up to an extra propagation of the control atom to the bottom left site before implementing $U^\mathrm{t}(\theta,0)$, as well as a propagation back to the original site after the tunneling gate implementation.
Upon displacing the control atom configuration by one site in the horizontal and/or vertical direction, the same procedure is repeated to implement the diagonal hopping process between all next-nearest-neighbor pairs.
Moreover, note that upon considering other specific control plane configurations, it is possible to synthesize arbitrary, long-range tunneling and interaction terms. Indeed, the dynamics of the extended Fermi-Hubbard model and variants thereof can be simulated in a fully digital way, see SM.

\clearpage
\appendix

\makeatletter
\@removefromreset{equation}{section}
\makeatother

\onecolumngrid
\section*{SUPPLEMENTAL MATERIAL }

\let\thesubsection\oldthesubsection
\setcounter{equation}{0}
\setcounter{figure}{0}
\setcounter{table}{0}
\setcounter{section}{0}
\setcounter{subsection}{0}
\renewcommand{\theequation}{S\arabic{equation}}
\renewcommand{\thefigure}{S\arabic{figure}}
\setlength\tabcolsep{10pt}
\setcounter{secnumdepth}{3}

\vspace{4mm}
\twocolumngrid

\section{Fermionic gate sequences}

Here we provide details on the control sequences discussed in the main text to realize the tunneling and interaction gate.

\subsection{Double-well Hilbert space}

Before describing in detail the gate sequences of the main text, we discuss our treatment of the double-well (DW) Hilbert space that appears in the scheme. 
We consider a single double well with two data modes (left and right), and similarly two control modes. The data modes can initially be in any state, while the control modes start with either no fermions or one fermion in the right well. The DW Hilbert space is spanned by the basis $\{\ket{n_{\mathrm{d,L}} n_{\mathrm{d,R}}}_\mathrm{d}\otimes\ket{n_{\mathrm{c,L}}n_{\mathrm{c,R}}}_\mathrm{c}\}$, with $n_{\sigma,j}$ denoting the number of $\sigma\in\{\mathrm{c},\mathrm{d}\}$ fermions in well $j\in\{\mathrm{L},\mathrm{R}\}$.

The DW Hilbert space $\mathcal{H}_{\rm DW}$ can thus be written as a tensor product of the Hilbert space of each mode, $\mathcal{H}_{\rm DW}=\mathcal{H}_{\mathrm d}\otimes\mathcal{H}_{\mathrm c}$. In the following, it is convenient to further divide each of these (four-dimensional) Hilbert spaces into a direct sum of (i) a two-level ``odd-parity" subspace, corresponding to a single fermion in the DW, i.e., spanned by $\ket{10}_\sigma$ and $\ket{01}_\sigma$, and (ii) a two-level ``even-parity" subspace, spanned by $\ket{00}_\sigma$ and $\ket{11}_\sigma$. 
Under the physical Hamiltonian Eq.~\eqref{eq:Hamiltonian} without the interaction $H_U$, the data and control modes in a DW evolve independently. In the even-parity subspace, tunneling is not possible, and thus $\ket{00}_\sigma$ and $\ket{11}_\sigma$ only acquire phases; on the other hand, the odd-parity subspace evolves as a driven two-level system (TLS). 
When the interaction is on, the d and c modes typically get entangled, making our tensor product decomposition impractical. However, this is not the case if tunneling is off and if one species (d or c) is in a specific number eigenstate before the interaction is applied: in that case, the interaction essentially acts as a potential gradient on the other species, conditioned on the state of the first one.
To make this explicit, take for example the Hamiltonian $H_U=U \sum_{j=\mathrm{L,R}} n_{\mathrm{d},j} n_{\mathrm{c},j}$, and assume that the data modes are in an arbitrary state $\ket{\psi}_\mathrm{d}$ while the control modes are in $\ket{01}_\mathrm{c}$. Then, the evolution of the entire state for time $t$ is 
\begin{align}
    &e^{-iH_U t} (\ket{\psi}_{\mathrm d} \otimes \ket{01}_{\mathrm c})\\
    &\qquad=(e^{-iUt n_{\mathrm{d,R}}}\ket{\psi}_{\mathrm d})\otimes\ket{01}_{\mathrm c},
\end{align}
which is equivalent to having a gradient on the data modes and keeping the control modes untouched. In particular, writing $\ket{\psi}_\mathrm{d}=\ket{\psi_\mathrm{e}}_{\mathrm d}+\ket{\psi_\mathrm{o}}_{\mathrm d}$ as a sum of even- and odd-parity subspace components, the odd-parity component $\ket{\psi_\mathrm{o}}$ undergoes a Bloch sphere precession of angle $Ut$. 
This discussion applies similarly to other basis states of the control modes, and also upon inverting the role of control and data fermions.

\subsection{Tunneling gate}
The tunneling gate should act on two nearest-neighbor data fermions as 
\begin{equation}
U^\mathrm{t}_{j,j+1}(\theta_1, \theta_2)=e^{-i\theta_1(e^{-i\theta_2}d^{\dagger}_j d_{j+1} +\rm H.c.)},
\end{equation}
while its action on the control fermions is irrelevant as long as they remain (after the full sequence) decoupled from the data fermions. For the protocols discussed here, the position of the control fermions is also preserved by the sequence.

The first step of the gate protocol is to adjust the lattice phases to split the lattice into DWs; with this, we can consider only two sites $j,j+1$ in a DW isolated from the rest of the system. The Hamiltonian in this DW is
\begin{align}
    H_{\rm DW}&=-J[d_\mathrm{R}^{\dagger}d_\mathrm{L}+c_\mathrm{R}^{\dagger}c_\mathrm{L}+\mathrm{H.c.}]\notag\\
    &\qquad+\sum_{\sigma=\mathrm{d,c}}\delta_{\sigma}n_{\sigma,\mathrm{L}}+U\sum_{i=\mathrm{L,R}}n_{\mathrm{d},i}n_{\mathrm{c},i},
\end{align}
where $J$ (tunneling), $U$ (interaction), and $\delta_\sigma$ (potential gradient, or ``detuning" in the TLS analogy) can be tuned freely.

Then, we apply the three-step global control sequence described in the main text. The first step \circledsub{1} has nonzero $J$ and $\delta_{\mathrm{d}}$, vanishing $\delta_{\mathrm{c}}$ and $U$, and duration $\tau=\pi/({2J})$. Since the interaction is off, the applied unitary can be written as $U_\mathrm{d}^{\circledsup{1}}\otimes U_\mathrm{c}^{\circledsup{1}}$, where in the odd-parity subspace $U_\mathrm{d}^{\circledsup{1}}$ and $U_\mathrm{c}^{\circledsup{1}}$ act as (having defined $A=\sqrt{1+(\delta_{\rm d}/J)^2 /4}$)
\begin{align}
    U_\mathrm{d}^{\circledsup{1},o}&=e^{-i\tfrac{\pi\delta_{\mathrm{d}}}{4J}}\left[\cos(\pi A)\mathbbm{1}-i\sin(\pi A) \frac{-2J\hat{\sigma}_{x,\mathrm{d}}+\delta_{\rm d} \hat{\sigma}_{z,\mathrm{d}}}{\sqrt{4J^2+\delta_{\mathrm{d}}^2}}\right], \\
    U_\mathrm{c}^{\circledsup{1},\mathrm{o}}&=i\hat{\sigma}_{x,\mathrm{c}},
\end{align}
and in the even-parity subspace they act as $U_\mathrm{d}^{\circledsup{1},\mathrm{e}}=e^{-i\pi\delta_{\rm d}/(2J) \, \ket{11}\bra{11}_\mathrm{d}}$ and $U_\mathrm{c}^{\circledsup{1},\mathrm{e}}=\mathbbm{1}$. Here, we define $\hat\sigma_{x,\sigma}=\ket{01}\bra{10}_\sigma + \ket{10}\bra{01}_\sigma$ and $\hat\sigma_{z,\sigma}=\ket{10}\bra{10}_\sigma - \ket{01}\bra{01}_\sigma$, for both $\sigma=\mathrm{d,c}$.
Note that at the start, the control modes can be only in the state $\ket{00}_{\mathrm{c}}$ or $\ket{01}_{\mathrm{c}}$. Thus, after this first step, they must end up in $\ket{00}_{\mathrm{c}}$ (in the first case) or $i\ket{10}_{\mathrm{c}}$ (in the second case).

The second step \circledsub{2} has vanishing $\delta_{\sigma}$ and $J$, and applies on-site interaction $U$ for a time $\tau$ such that $U\tau=2\theta_1$. If there is a control fermion, it must be in the left well (due to the action of the first step, see above). Therefore, this second step acts as $\mathbbm{1}_{\mathrm c} \otimes U_{\mathrm{d}}^{\circledsup{2}}$: if there is no control fermion $U_{\mathrm{d}}^{\circledsup{2}}=\mathbbm{1}$, while if there is one $U_{\mathrm{d}}^{\circledsup{2}}=e^{-2i \theta_1 n_{\mathrm{d,L}}}$. In the latter case, we distinguish between the odd-parity subspace, where, using the relation $n_\mathrm{\sigma,L}=(1+\hat\sigma_{z,\sigma})/2$, we have 
\begin{align}
    U_{\mathrm{d}}^{\circledsup{2},\mathrm{o}}=e^{-i\theta_1} e^{-i\theta_1 \hat{\sigma}_{z,\mathrm{d}}},
\end{align}
and the even parity one, where $U_{\mathrm{d}}^{\circledsup{2},\mathrm{e}}=e^{-2i\theta_1\ket{11}\bra{11}_\mathrm{d}}$.

Finally, the third step \circledsub{3} is the inverse of the first one $(J, \delta_\mathrm{d})\rightarrow (-J, -\delta_\mathrm{d})$. While inverting $\delta_\mathrm{d}$ can be straightforwardly realized in an experimental setup, the inversion of the sign of $J$ can for instance be realized by switching on potential gradients ($\delta_\sigma$) before and after, i.e., $e^{-iH_{\mathrm{DW}}\tau}\rightarrow e^{-i\pi \sum_\sigma n_{\sigma,\mathrm{L}}}e^{-iH_{\mathrm{DW}}\tau}e^{i\pi \sum_\sigma n_{\sigma,\mathrm{L}}}$. 
Step \circledsub{3} can again be expressed as a tensor product of unitaries, with the factors being the inverses of the ones from the first steps $U_\sigma^{\circledsup{3}}=(U_\sigma^{\circledsup{1}})^\dagger$.

Knowing the action of the three individual steps, we can now derive that of the sequence on the data modes ($U_\mathrm{d}^\mathrm{tot}$), in both the ``with control'' and ``without control'' case. If there is no control, the action of the sequence is straightforward: the second step applies the identity, thus the first and third step cancel each other out, and $U_\mathrm{d}^\mathrm{tot}=\mathbbm{1}$. If there is a control, its state is unchanged ($U_\mathrm{c}^\mathrm{tot}\ket{01}_\mathrm{c}=\ket{01}_\mathrm{c}$), while the data modes evolve nontrivially. In the even-parity subspace, we have $U_\mathrm{d}^{\mathrm{tot},\mathrm{e}}=e^{-i2\theta_1\ket{11}\bra{11}_\mathrm{d}}$, and in the odd-parity one, 
\begin{align}
    U_\mathrm{d}^{\mathrm{tot},\mathrm{o}}&=e^{-i\theta_1}\; ({U_\mathrm{d}^{\circledsup{1},\mathrm{o}}})^\dagger e^{-i\theta_1 \hat{\sigma}_{z,\mathrm{d}}}U_\mathrm{d}^{\circledsup{1},\mathrm{o}}\\
    &=e^{-i\theta_1}\; \exp\left[-i\theta_1{(U_\mathrm{d}^{\circledsup{1},\mathrm{o}})}^\dagger \hat{\sigma}_{z,\mathrm{d}} U_\mathrm{d}^{\circledsup{1},\mathrm{o}}\right].
\end{align}
The operator $(U_\mathrm{d}^{\circledsup{1},\mathrm{o}})^\dagger \hat{\sigma}_{z,\mathrm{d}} U_\mathrm{d}^{\circledsup{1},\mathrm{o}}$ is a weighted sum of Pauli operators $\hat{\sigma}_{x,\mathrm{d}}$, $\hat{\sigma}_{z,\mathrm{d}}$, and $\hat{\sigma}_{y,\mathrm{d}}=i\ket{01}\bra{10}_\mathrm{d} -i \ket{10}\bra{01}_\mathrm{d}$. 
For this to implement the tunneling gate defined in the main text, we want the coefficient of $\hat{\sigma}_{z,\mathrm{d}}$ to vanish: this implies 
\begin{align}
    \delta_\mathrm{d}^2+4J^2\cos\left(\pi A\right)=0,
\end{align}
which is satisfied for $\delta_\mathrm{d}/J=\pm 1.5974$ (we assume $+$ from now on). Consequently, since $\hat{\sigma}_{x,\mathrm{d}}$ and $\hat{\sigma}_{y,\mathrm{d}}$ act respectively as $d^\dagger_\mathrm{L} d_\mathrm{R}+\mathrm{H.c.}$ and $-id^\dagger_\mathrm{L} d_\mathrm{R}+\mathrm{H.c.}$, the total unitary on data fermions (now in the entire 4-dimensional data Hilbert space) can be rewritten as

\begin{align}
    U_\mathrm{d}^\mathrm{tot}= e^{-i\theta_1(n_\mathrm{d,L}+n_\mathrm{d,R})} \exp[-i\theta_1(e^{-i\alpha}d_{\mathrm{L}}^\dagger d_{\mathrm{R}}+\mathrm{H.c.})],
\end{align}
where $e^{-i\alpha}$ is given by the coefficients of $\hat{\sigma}_{x,\mathrm{d}}$ and $\hat{\sigma}_{y,\mathrm{d}}$ as
\begin{align}
    e^{-i\alpha}=-J\frac{4\delta_\mathrm{d} \sin^2(\tfrac{\pi A}{2})-2i\sqrt{4J^2+\delta_\mathrm{d}^2}\sin(\pi A)}{4J^2+\delta_\mathrm{d}^2},
\end{align}
i.e., $\alpha\approx 2.4959$ for the chosen ratio $\delta_\mathrm{d}/J$. This unitary, up to a single-mode $\theta_1$ phase, implements the tunneling gate $U^\mathrm{t}_{j,j+1}(\theta_1,\theta_2=\alpha)$. To implement the tunneling gate with arbitrary $\theta_2$, we can switch on potential gradients before and after the gate sequence, i.e.,
\begin{align}
    &e^{i(\alpha-\theta_2) n_\mathrm{d,L}}U_\mathrm{d}^\mathrm{tot} e^{i(\theta_2-\alpha) n_\mathrm{d,L}}\notag\\
    &\qquad=e^{-i\theta_1(n_\mathrm{d,L}+n_\mathrm{d,R})} \exp[-i\theta_1(e^{-i\theta_2}d_{\mathrm{L}}^\dagger d_{\mathrm{R}}+\mathrm{H.c.})].
\end{align}
While, without loss of generality, we have assumed an initial placement of the control fermion on the right well R, the same gate sequence applies to the other placement on well L (up to implementing the tunneling operation $\propto d_\mathrm{L}^\dagger d_\mathrm{R}$ with the opposite $-\theta_1$).

\subsection{Interaction gate}

The last gate of our gate set is the interaction gate, which acts on the fermions as
\begin{align}
    U_{j,j+1}^\mathrm{int}(\theta)=e^{-i\theta n_{\mathrm{d},j} n_{\mathrm{d},j+1}},
\end{align}
and keeps the control fermion in the right well (if there is one).

The first step of our interaction gate protocol is to apply the tunneling gate sequence with tunneling angle $\theta_1=\pi/2$ (see previous subsection) with the roles of the control and data fermions inverted: this is realized by exchanging the data and control gradients in the tunneling control sequence, i.e., $\delta_\mathrm{d}=0$ and $\delta_\mathrm{c}/J=1.5974$. 
Here we assume for simplicity $\theta_2=0$, without loss of generality. 
Following and extending the derivation above, this thus keeps the data fermions unaffected and acts on the control fermions as [see also the End Matter (EM)]:
$U_\mathrm{c}=\mathbbm{1}$ (for data in $\ket{00}_\mathrm{d}$), $U_\mathrm{c}=e^{-i(\pi/2)(n_\mathrm{c,L}+n_\mathrm{c,R})} e^{-i(c_{\mathrm{L}}^\dagger c_{\mathrm{R}}+c_{\mathrm{R}}^\dagger c_{\mathrm{L}})\pi/2}$ (for data in $\ket{01}_\mathrm{d}$), $U_\mathrm{c}=e^{-i(\pi/2)(n_\mathrm{c,L}+n_\mathrm{c,R})} e^{+i(c_{\mathrm{L}}^\dagger c_{\mathrm{R}}+c_{\mathrm{R}}^\dagger c_{\mathrm{L}})\pi/2}$ (for data in $\ket{10}_\mathrm{d}$), or $U_\mathrm{c}=e^{-i\pi(n_\mathrm{c,L}+n_\mathrm{c,R})}$ (for data in $\ket{11}_\mathrm{d}$).

The second step is to switch on an energy offset $\delta_\mathrm{c}$ for a time $\tau=-\theta/(2\delta_\mathrm{c})$. For odd-parity data basis states, the first step has made the control atom move to the left well, and thus this second step imprints a phase $\theta/2$. Conversely, for even-parity data states, the control has stayed in the right well and no phase is imprinted.

The third step consists in repeating the first one. Consequently, for odd-parity data configurations the control atom returns to its original position, after having acquired a phase, while in the even-parity data configurations it stays in its original location. The total unitary describing the evolution of the control atom is given, for all data configurations, by:
\begin{widetext}
\begin{align}
U_\mathrm{c}^\mathrm{tot}=
\begin{cases}
    e^{-i\theta n_\mathrm{c,L}/2}  &\mathrm{for}\; \ket{00}_\mathrm{d},\notag\\
    e^{-i\pi(n_\mathrm{c,L}+n_\mathrm{c,R})} e^{-i(c_{\mathrm{L}}^\dagger c_{\mathrm{R}}+c_{\mathrm{R}}^\dagger c_{\mathrm{L}})\pi/2} \notag  e^{i\theta n_\mathrm{c,L}/2} e^{-i(c_{\mathrm{L}}^\dagger c_{\mathrm{R}}+c_{\mathrm{R}}^\dagger c_{\mathrm{L}})\pi/2} & \mathrm{for}\; \ket{01}_\mathrm{d},\notag\\
    e^{-i\pi(n_\mathrm{c,L}+n_\mathrm{c,R})} e^{+i(c_{\mathrm{L}}^\dagger c_{\mathrm{R}}+c_{\mathrm{R}}^\dagger c_{\mathrm{L}})\pi/2} e^{i\theta n_\mathrm{c,L}/2} e^{+i(c_{\mathrm{L}}^\dagger c_{\mathrm{R}}+c_{\mathrm{R}}^\dagger c_{\mathrm{L}})\pi/2} &\mathrm{for}\; \ket{10}_\mathrm{d},\notag\\
    e^{-i\theta n_\mathrm{c,L}/2}e^{-2i\pi(n_\mathrm{c,L}+n_\mathrm{c,R})} &\mathrm{for}\; \ket{11}_\mathrm{d}.        
\end{cases}
\end{align}
\end{widetext}

This can be further expanded to compute the effect of the sequence on all $\{\ket{n_{\mathrm{d,L}} n_{\mathrm{d,R}}}_\mathrm{d}\otimes\ket{n_{\mathrm{c,L}}n_{\mathrm{c,R}}}_\mathrm{c}\}$ basis states. From that, it appears that the $\ket{00}_\mathrm{c}$ and $\ket{01}_\mathrm{c}$ control states are preserved, and correspond to different unitaries applied to the data modes. For a double-well with no control atom ($\ket{00}_\mathrm{c}$), the sequence applies the identity to the data Hilbert space. For a double-well containing a control atom ($\ket{01}_\mathrm{c}$), the system picks up a phase factor $e^{i\theta/2}$ for data in $\ket{01}_\mathrm{d}$ or $\ket{10}_\mathrm{d}$, and no phase otherwise. This translates to a unitary acting on the data modes as 
\begin{align}
    U_{\mathrm{d}}^\mathrm{tot}=e^{-i\theta (2n_\mathrm{d,L} n_\mathrm{d,R}-n_\mathrm{d,L} - n_\mathrm{d,R})/2},
\end{align}
which implements the desired interaction gate up to single-mode phases.

\section{Alternative operation mode: the conveyor belt}

In the main text we introduced a framework to realize a 1D fermionic quantum circuit in a 1D uniformly controlled optical lattice.
There exist alternative operation modes to achieve the same goal, unlocking additional features at the cost of some disadvantages; we now describe one, the ``conveyor belt" approach, inspired by similar ideas in Rydberg atom arrays~\cite{cesa2023universal} and superconducting circuits ~\cite{riccardi2024global}.

\begin{figure}[t]
    \centering
    \includegraphics[width=\linewidth]{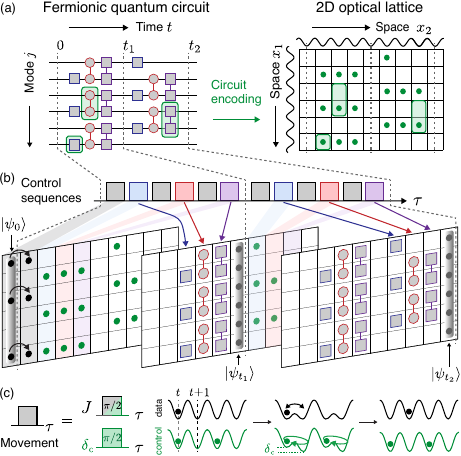}
    \caption{\textbf{Conveyor-belt operation mode.} (a) A fermionic quantum circuit, featuring a periodic sequence of phase, tunneling and interaction gate-layers. The 1D quantum circuit is realized by encoding the space-time location $(j,t)$ of the gates in the location $(x_1, x_2)$ of control fermions (green dots) in a 2D optical lattice. 
    (b) The quantum state is instead encoded in the data fermion (black dots) configuration, aligned along a 1D column.
    Similar to the standard approach, the circuit is applied by alternating data fermion movement with tailored global control sequences acting non-trivially on data atoms only where the control fermion sits. Different columns of control fermions in the 2D lattice corresponds to different circuit layers. (c) A key element for this framework is the selective movement of the data atoms. While control fermions are fixed at a given circuit time $t\; (\equiv x_2)$, the data atom is transported as follows: First, the lattice is rearranged into double wells, such that the data atom sits on a left well~L. We then allow tunneling with strength $J$ for time $\tau=\pi/(2J)$, in presence of a gradient $\delta_\mathrm{c}$ for control fermions, whose value is chosen such that control particles return to their original wells in each step. Therefore, the data atom proceeds to the next well $t+1$, while control atoms remain localized.}
    \label{fig:SMFig1}
\end{figure}

Here, we consider a 2D uniformly controlled optical lattice (OL), and we identify the space and time direction ($j, t$) of the quantum circuit with the vertical and horizontal OL direction ($x_1, x_2$), see Fig.~\ref{fig:SMFig1}(a). 
The circuit is then encoded by placing control fermions in the 2D lattice according to the space- and time-location of the quantum gates. The 1D quantum state is instead encoded in the data atoms, placed in a single column, the \textit{data} column, along the $x_1$ direction. The initial atom placement in localized wells is the only step in this framework requiring some degree of local programmability. 

We then apply the 1D fermionic circuit using the control sequences for the three different gate types discussed in the main text. The steps have a net non-trivial effect only on the DWs hosting control heads and located in the data column.
Contrary to the standard approach, in the conveyor-belt operation mode we move the data column in circuit-time direction $t\; (\equiv x_2)$, while the control particles remain static, see Fig.~\ref{fig:SMFig1}(c). This selective movement can be realized with the gate discussed in Fig.~\ref{fig:Fig1}(c) and End Matter (EM) upon exchanging the role of control and data atoms, and realizing a DWs structure in $x_2$-direction. The movement sequence can be simplified to a single tunneling step with duration  $\tau=\pi/(2J)$ and gradient $\delta_\mathrm{c}=2J\sqrt{4m^2-1}$ ($m\in \mathbb{N}$), see Fig.~\ref{fig:SMFig1}(c). In this case, the extra single-mode phases are acquired by the control fermions. After the movement, they are localized in a single site and thus are in a product state with data atoms: therefore, we can forget about the extra phases, without the need of a second tunneling step with opposite gradient.

We now compare the features of the two operation modes, starting from the number of control heads necessary for the most general circuit. While in the standard approach we require a single control fermion, here we need as many particles as the number of gates. Moreover, the conveyor-belt approach requires an higher degree of local programmability in the initial circuit encoding, since multiple control atoms need to be precisely placed at specific sites in the 2D lattice without any regular pattern. On the contrary, in the standard approach the control head can be initially located anywhere in the 1D OL, and then moved to the desired site.\\
\noindent
Another difference between the two operation modes regards the parallelization.
The conveyor-belt setup is particularly useful for circuits featuring a periodic alternation of different gate-type layers; specifically, we consider cases in which the gates within each layer are performed always with a fixed angle $\{ \theta^\mathrm{p}, \theta_1^\mathrm{t}, \theta_2^\mathrm{t}, \theta^\mathrm{int} \}$, but at varying locations of the 1D quantum state, see, e.g., the circuit in Fig.~\ref{fig:SMFig1}(a). Under these conditions, we can simultaneously prepare multiple copies of the same initial quantum state at different circuit times. 
More precisely, we prepare multiple copies of a given initial product state, separated by the same number of lattice sites as the control heads encoding two gate layers of the same type. All the copies initially lay in an initialization region, distinct from the processing region where the control heads have been placed. We then run the quantum circuit by alternating between movement and processing gates: in particular, after a number of movement steps equal to the above spacing, a new copy enters the processing region and we start applying the circuit on it. Since at a given physical time, all the data columns in the processing region overlap with control heads encoding circuit layers of the same gate-type, using a single global control sequence it is possible to implement a different circuit layer on each copy.
In this way, we can prepare multiple copies at \textit{different} circuit times.
This capability enables the calculation of otherwise experimentally challenging quantities, such as out-of-time-order correlators~\cite{bohrdt2017scrambling,swingle2018unscrambling} and Loschmidt echos~\cite{cavallar2025phasesensitive}.
On the contrary, the standard framework allows us to efficiently prepare multiple copies at the \textit{same} circuit time; this is useful to extract many-body observables~\cite{pichler2016measurement,lukin2019probing}, in contrast to the usual randomized protocols~\cite{naldesi2023fermionic}.

\section{Hybrid analog-digital operation mode with two spin species}

In this Appendix, we illustrate a method to implement the hybrid analog-digital Trotter evolution described in the main text using only two fermionic species. To this end, we consider a two-layer setup analogous to the one of the main text [see Fig.~\ref{fig:Fig3}(b)]. Now, however, the ``control" plane is filled with fermions of one of the two ``data" species (say $\uparrow$). 
We further require the capability to drive the $\{\uparrow,\downarrow\} $ transition with a microwave (MW) spin-flip Hamiltonian
\begin{equation}
    H_\mathrm{MW}=\frac{\Omega}{2}\sum_j (f_{\uparrow, j}^\dagger f_{\downarrow, j} + f_{\downarrow, j}^\dagger f_{\uparrow, j}),
\end{equation}
with Rabi frequency~$\Omega$.

We first describe how we implement the digital operations. The key idea is to alternatively use each species as control particles for the other one. To this end, we first use our spin-selective movement protocol to move all $\downarrow$-fermions from the data plane to the control one, while keeping the $\uparrow$-fermions untouched. This requires a spin-dependent gradient in the vertical direction, i.e., between the two planes. Then, we apply the series of gate sequences required to implement the desired unitary on $\downarrow$-fermions. Again, the sequences do not affect the remaining $\uparrow$-particles in the data plane.  
Finally, after moving the $\downarrow$-fermions back to the data plane, we apply a MW $\pi$-pulse (i.e., of duration $\tau=\pi/\Omega$) on the control plane, which flips the $\uparrow$-fermions in this plane to $\downarrow$. This plane-specific pulse can be realized in a globally controlled way by splitting it into two $\pi/2$-pulses and 
applying, before and after the second pulse, a spin-dependent potential gradient between the data and control planes. Specifically, we choose, in the Hamiltonian $H_\delta$ of Eq.~\eqref{eq:HamDetuning}, $j=0,1$ to correspond respectively to the control plane and the data plane, and take $(\delta_\uparrow-\delta_\downarrow) \tau= \pi$. This ensures that fermions in the data plane return to their initial spin state, while fermions in the control plane change state.
The whole sequence is then repeated, exchanging the role of the two spin species, to implement the desired digital operation also on $\uparrow$ data fermions.

For the analog evolution, the challenge resides in ensuring that the fermions in the control plane do not evolve, such that the pattern of control fermions remains intact for the following Trotter step. During the analog evolution, there are no data fermions in the control plane, and the evolution of control fermions is limited to free particle tunneling, i.e., $U_{\tau} = e^{-iH_{J\sigma} \tau}$, where $H_{J\sigma}=-J\sum_j (f_{\sigma,j}^\dagger f_{\sigma,j+1} + \mathrm{H.c.})$ is the tunneling Hamiltonian between neighboring sites for the spin species $\sigma$ of the control plane. In the following, we describe how a global control sequence allows us to inhibit this unwanted tunneling process. For a Trotter step of duration $\tau$, this is achieved by alternating the analog evolution for a time $\tau/2$ with the application of additional phases in the control plane. Specifically, by applying $\pi$ phases via the unitary $U_\pi~=~\prod_{i\in \mathrm{CB}}\exp(i \pi n_{\sigma,i} )$ to the control plane, where CB denotes mode indices corresponding to a checkerboard pattern, we invert the sign of the tunneling process as $ U_\pi U_{\tau/2} U_\pi^\dagger = (U_{\tau/2})^\dagger$. Therefore, the resulting evolution is given by $U_{\tau/2} U_\pi U_{\tau/2} U_\pi^\dagger = \mathbbm{1}$ and the control plane does not evolve during the analog evolution step. To obtain the correct evolution in the data plane, however, the data fermions need to remain unaffected by the additional phases.
To this end, we synthesize the checkerboard $\pi$-phase $U_\pi$ for the control plane in the following way (w.l.o.g. we again consider the case where fermions in the control plane are initially in spin state $\sigma=\uparrow$).
We first apply a spin-dependent gradient $\delta_\uparrow$ in both spatial directions, for time $\tau=\pi/\delta_\uparrow$, which applies a checkerboard $\pi$-phase to $\uparrow$-fermions in both planes. We then flip control-plane fermions from $\uparrow$ to $\downarrow$ following the procedure detailed above for the digital part. At this point, applying again the gradient for a time $\tau=\pi/\delta_\uparrow$ cancels the phases acquired in the data plane (all phases add to multiples of $2\pi$), while having no effect on the control plane, where the $\pi$-phase in checkerboard pattern persists. Finally, we flip control-plane fermions back to $\uparrow$. In total, only fermions in the control plane have acquired phases, thus implementing $U_\pi$: this sequence can be therefore used to selectively undo the analog evolution of the control atoms.

\section{Fully digital 2D Trotter evolution}
In the main text we discussed an hybrid analog-digital approach to simulate a 2D long-range, spinful Fermi-Hubbard model.
In this Appendix, we show how the same result can be achieved in a fully digital way.

\begin{figure}[t]
    \centering
    \includegraphics[width=0.85\linewidth]{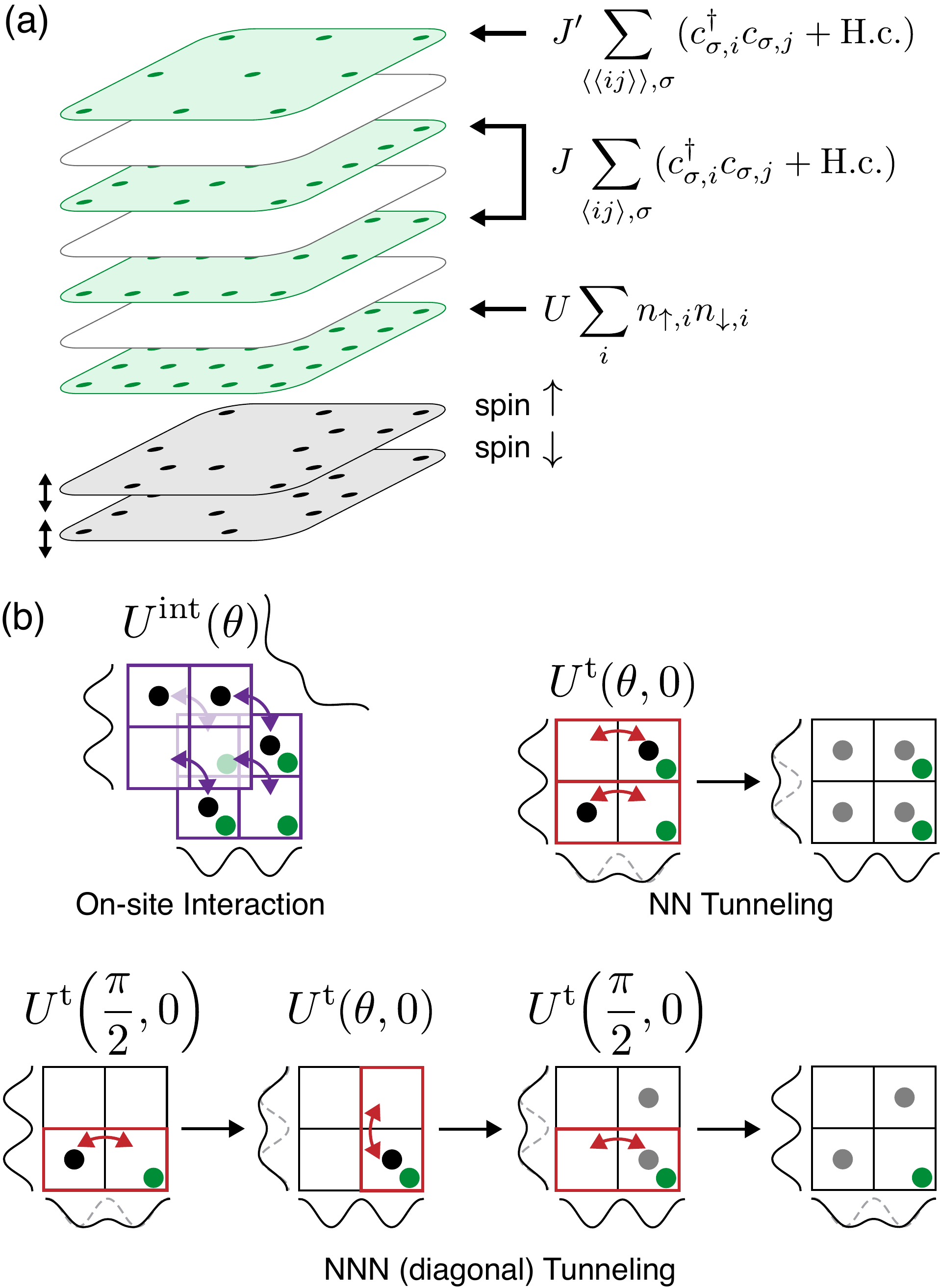}
    \caption{\textbf{Fully digital Trotter evolution.} (a) We simulate the dynamics of an extended 2D spinful Fermi-Hubbard model Eq.~\eqref{eq:SMextendedFH}, featuring long-range tunneling terms, in a fully digital way. We encode the two spin species into two different data fermion planes (grey planes with black dots). We also prepare 2D planes loaded with control fermions (green planes with green dots), arranged in highly symmetric configurations: each of these configurations digitally implements a different term of the Hamiltonian. (b) Operation sequence to implement the evolution under the different terms in Eq.~\eqref{eq:SMextendedFH}, via an alternating sequence of lattice rearrangements and the digital gates discussed in the main text.
    }
    \label{fig:SM_digital}
\end{figure}

Specifically, we want to investigate the dynamics of the Hamiltonian
\begin{align}
     H_{\mathrm{FH+}} &= \textstyle H_{\mathrm{FH}} + H_\mathrm{LR},   \label{eq:SMextendedFH} \\
     H_{\mathrm{FH}} & = \textstyle -\sum_{\langle ij\rangle,\sigma} J(c^\dagger_{\sigma,i}c_{\sigma,j} + \mathrm{H.c.}) + U\sum_i n_{\uparrow,i}n_{\downarrow,i} \nonumber ,\\
     H_\mathrm{LR} &= - \sum_{\llangle ij\rrangle,\sigma} J'(c^\dagger_{\sigma,i}c_{\sigma,j} + \mathrm{H.c.})\nonumber;   
\end{align}
see the main text for more details.
The total evolution unitary $U(t)=\mathrm{exp}{(-i H_\mathrm{FH+}t)}$ can be then Trotterized as $U(t)\approx (\prod_\alpha U_{\Delta t}^\alpha)^{t/\Delta t} = (\prod_\alpha e^{-i H_\alpha \Delta t})^{t/\Delta t}$ with  $\Delta t$ the Trotter step and $H_\alpha$ corresponding to the different terms in the Hamiltonian.
The Fermi-Hubbard Hamiltonian acts on two different spins $\sigma=\uparrow, \downarrow$: we can encode them either in two different atomic species placed in a single 2D plane (as we do in the hybrid approach), or use a single atomic species in two planes, with each plane simulating a different spin, see Fig.~\ref{fig:SM_digital}. In the following, we consider the latter encoding; in our architecture the two data planes are then moved simultaneously via the movement sequence.
We further prepare 2D planes loaded with control fermions and exhibiting specific configuration patterns, each allowing the implementation of a different unitary $\{e^{-i H_\alpha \Delta t}\}_\alpha$.
Both data and control planes are stacked vertically in a 3D optical lattice, see Fig.~\ref{fig:SM_digital}.

The scheme for the digital implementation of the Trotterized evolution consists of two steps: First, we overlap one data plane with the desired control plane. This can be achieved by applying either the control [see Fig.~\ref{fig:Fig1}(c)] or the data [see Fig.~\ref{fig:SMFig1}(c)] movement sequence in vertical direction. Then, using a combination of movement sequences, as well as tunneling and interaction gates we implement the unitary evolution gates $U_{\Delta t}^\alpha$. We repeat this step for the second data species and for each term in the Hamiltonian.
In the following, we discuss in details the specific sequences.

We start by the on-site interaction $U n_{\uparrow, i}n_{\downarrow,i}$, which requires a fully-filled control plane. After overlapping it with one data plane, we realize a DW structure between the two data planes.
At this point we simply apply the global control sequence for the density-interaction term with $\theta = U\Delta t$.\\ \noindent
For the implementation of the tunneling term in the horizontal resp. vertical direction of a given data plane, we instead require a 2D control configuration, where every second column resp. row is fully filled. After overlapping the control plane with a data plane, the evolution unitary is implemented by performing the global control sequence for the tunneling gate with $\theta_1=-J\Delta t, \theta_2=0$.
Alternatively, the nearest-neighbor tunneling can be realized with a single control plane in a checkerboard configuration, at the cost of having control fermions in either the left or the right site of the DW: as discussed above, a control atom in the left well implements the tunneling gate with opposite angle $-\theta_1$. To correct for this, we can engineer an echo sequence based on phase gates with angle $\theta=\pi/2$.
\\ \noindent
The last term we simulate is the next-nearest-neighbor tunneling: in the corresponding control plane only modes with even row and column indices are filled. 
The evolution unitary is then implemented as follows. We start with a DW configuration in the horizontal direction, and we apply a tunneling gate $U^\mathrm{t}(\pi/2,0)$, conditional on the presence of a control fermion in the DW, thus effectively exchanging the two data modes. We then isolate the system into DWs along the vertical direction, and apply the control sequence for the tunneling gate $U^\mathrm{t}(-J'\Delta t,0)$: this implements the unitary  $e^{-i H_\alpha \Delta t}$, with $H_\alpha=H_\mathrm{LR}$. 
Finally, we rearrange the lattice to exhibit DWs in the horizontal direction, and apply $U^\mathrm{t}(\pi/2,0)$ again.
Note that to apply the diagonal term in the opposite direction, i.e., from top left to bottom right, we need to propagate the control atoms to the left site of the DW prior to the lattice rearrangement.
\\ \noindent
For all non-local terms, the corresponding sequence has to be repeated upon translating the control plane by one site in horizontal and/or vertical direction with respect to the data plane to make sure that the unitary evolution gate is applied on all possible mode pairs.

An extension to Hamiltonians featuring even longer-range tunneling terms (between modes $i$ and $j$) is straightforward. In this case, the key step is the conditional tunneling gate $U^\mathrm{t}(\pi/2,0)$ for data fermions, i.e., a fermionic SWAP (fSWAP) operation, as well as the movement of control fermions. By alternating them with lattice rearrangements, we can move a data fermion originally at mode $i$ to the other well of the DW hosting mode $j$. We then locally apply the tunneling gate $U^\mathrm{t}(-\tilde{J}\Delta t, 0)$, thus effectively implementing the long-range tunneling term $\tilde{J}(d_i^\dagger d_j + \mathrm{H.c.})$. In an analogous way, we can introduce long-range interspecies interaction terms: this requires to conditionally move the data fermion at mode $i$ to site $j$, and then realize a DW structure between the two data planes to enable the implementation of the density-interaction gate, in analogy to what we propose for the on-site interaction. 

Note that the alternation of fSWAP, control propagation gates, and lattice rearrangements requires a number of steps linear in the distance between the targeted modes: an alternative approach involves loading the data fermions to an additional plane, shuttling them to the desired site, and moving them back to the original data plane. 

\end{document}